\documentclass[aip,rsi,amsmath,reprint,amssymb]{revtex4-1}
\usepackage{graphicx}
\usepackage{textcomp}
\usepackage{mathptmx}
\usepackage{hyperref}

\begin{document}
\thispagestyle{empty}
\onecolumngrid

\vspace*{5cm}
\begin{center}
\huge
The following article appeared in\\\vspace{6pt} \mbox{Rev. Sci. Instrum. 85, 014702 (2014)}\\\vspace{6pt} and may be found at\\\vspace{6pt} \url{http://dx.doi.org/10.1063/1.4856475}.\\
\vspace{4cm}
Copyright 2014 American Institute of Physics.\\ This article may be downloaded for personal use only.\\ Any other use requires prior permission of the author and the American Institute of Physics.
\end{center}
\cleardoublepage

\title{Surface-resistance measurements using superconducting stripline resonators} 

\preprint{report number}

\author{Daniel Hafner}
\noaffiliation
\author{Martin Dressel}
\noaffiliation
\author{Marc Scheffler}
\email[]{scheffl@pi1.physik.uni-stuttgart.de}
\noaffiliation

\affiliation{1. Physikalisches Institut, Universit\"at Stuttgart, D-70550 Stuttgart, Germany}

\begin{abstract}
We present a method to measure the absolute surface resistance of conductive samples at a set of GHz frequencies with superconducting lead stripline resonators at temperatures 1~- 6~K. The stripline structure can easily be applied for bulk samples and allows direct calculation of the surface resistance without the requirement of additional calibration measurements or sample reference points. We further describe a correction method to reduce experimental background on high-Q resonance modes by exploiting TEM-properties of the external cabling. We then show applications of this method to the reference materials gold, tantalum, and tin, which include the anomalous skin effect and conventional superconductivity. Furthermore, we extract the complex optical conductivity for an all-lead stripline resonator to find a coherence peak and the superconducting gap of lead.
\end{abstract}

\setcounter{page}{1}
\maketitle

\section{introduction}
Microwave spectroscopy on a solid, i.e.\ measuring its frequency-dependent response at GHz frequencies, gives experimental access to the optical properties of a material at comparably low frequencies (30~GHz $\approx$ 1~cm$^{-1}$ $\approx$ 124~$\mu$eV $\approx$ 1.4~K).\cite{Dressel2002a} We are interested in the study of conductive bulk materials such as metals or superconductors; here microwave spectroscopy can reveal low-energy charge dynamics and electronic or magnetic excitations.\cite{Dressel2006,Basov2011,Scheffler2013} These low probing energies correspond to the low temperatures where many of the most interesting effects occur in this context. Therefore, several experimental approaches have been developed to study the microwave properties of metals and superconductors at cryogenic temperatures.
Ideally, the technique for such an experiment would have the following properties: broad spectral range (complete frequency dependence), full complex response (amplitude and phase), good sensitivity (also applicable for low-loss samples), and compatibility with cryogenic temperatures.

Unfortunately, so far there is no technique available that fulfills all of these options:
Corbino reflection spectroscopy is applicable in extremely broad frequency ranges (up to 40 GHz)\cite{Wu1995,Scheffler2007} and at cryogenic temperatures,\cite{Booth1994,Stutzman2000,Scheffler2005a,Kitano2008} is phase-sensitive, and has been used in the study of conventional and exotic superconductors \cite{Booth1996a,Ohashi2006,Steinberg2008,Pompeo2010,Liu2011}, correlated metals,\cite{Scheffler2005c,Scheffler2006,Scheffler2010}, and materials close to a transition between conductive and insulating states.\cite{Lee2001} 
While the Corbino approach is applicable for bulk samples of lossy conductors,\cite{Lee2001,Schwartz2000a,Scheffler2005a} its comparably poor sensitivity restricts its applicability for highly conductive metals and superconductors to geometrically strongly confined samples,\cite{Booth1994,Scheffler2005a,Scheffler2007} such as thin films. Since many particularly interesting conducting materials are only available as bulk samples, the Corbino approach is not applicable here.
A broadband microwave technique that is sensitive enough to study bulk superconductors well below the critical temperature is the bolometric spectrometer,\cite{Turner2003,Turner2004,Bobowski2010} but this only measures the amplitude of the microwave absorption, i.e.\ it does not give phase information.

An approach that can both be very sensitive and reveal phase information are resonant measurements, where the sample acts as perturbation to a microwave resonator. From measurements of resonance frequency and quality factor, one can in principle determine amplitude and phase of the microwave properties of the sample. Another advantage are the rather loose requirements concerning calibration of microwave lines that transmit the signal in the cryogenic environment. Amongst the different resonator techniques,\cite{Deri1986,Bonn1991} cavity perturbation has been used for several decades already and probably is the most generically applied microwave technique to study metals and and in particular superconductors at cryogenic temperatures.\cite{Klein1993,Wu1993,Shibauchi1994,Dressel1994,Peligrad1998,Hosseini1999,Hashimoto2009} However, there are two main disadvantages of cavity measurements: firstly, it is often very difficult to obtain absolute values. Secondly, and more fundamental, cavity resonators are usually operated at a single resonance frequency, and therefore one does not obtain any information on the frequency dependence. Obtaining spectral information from cavity measurements either requires operation of a cavity at different modes or of separate cavities;\cite{Hosseini1999} in both cases it is difficult to obtain quantitative information. 
A recent development to use the high sensitivity of a resonator and simultaneously obtain spectral information is operating a dielectric resonator at different resonance frequencies, and this technique has been applied successfully to the study of bulk superconductors at mK temperatures.\cite{Huttema2006,Truncik2012}

In our work, we follow a somewhat different experimental approach. Our goal are cryogenic microwave measurements that are sensitive enough for bulk samples of good metals or superconductors, that reveal the frequency dependence, and that are compatible with operation at mK temperatures in a dilution refrigerator. All these requirements are met by superconducting stripline resonators where the sample acts as one of the ground planes.
Superconducting stripline (also called triplate) resonators,\cite{DiIorio1988,Oates1990,Nguyen1993,Revenaz1994,Belk1996} as well as the similar microstrip resonators,\cite{Anlage1989,Anlage1991, Andreone1993,Andreone1994} have been used since the 1980s to study thin films of superconductors. A related technique are superconducting coplanar resonators, which have also been used to study the microwave properties of superconducting thin films,\cite{Rauch1993,Song2009} and which recently became very popular for photon detection (as microwave kinetic inductance detectors)\cite{Day2003,Zmuidzinas2012} and in the field of quantum information science (e.g.\ as building blocks for circuit QED architecture).\cite{Wallraff2004,Goeppl2008} While coplanar resonators have advantages compared to stripline and microwave resonators in terms of achievable quality factors, scalability, and ease of fabrication, it is difficult to combine them with bulk samples if one is interested in quantitative information about the sample's microwave properties.\cite{Scheffler2013}

Therefore, we employ superconducting stripline resonators.\cite{Scheffler2012} While previous applications of this technique focused on thin film samples,\cite{DiIorio1988,Oates1990,Nguyen1993,Revenaz1994,Belk1996} DiIorio \textit{et al.} also demonstrated measurements on a bulk sample of gold as a reference.\cite{DiIorio1988} Starting from these previous works, we have developed stripline resonators for the particular goal of studying bulk samples of highly conductive metals or superconductors. With the high achievable quality factors of these resonators, we are sensitive enough to detect the response of such low-loss samples. Furthermore, because these resonators are based on one-dimensional transmission lines, we can easily operate them not only at the fundamental mode, but also at numerous higher harmonics. Thus we obtain data at a set of roughly equidistant frequencies, with typical frequency spacings around 1~GHz. While this cannot compete with the complete frequency dependence accessible with broadband techniques, it is sufficient for many questions in the field of superconductors and correlated metals, where spectral features are usually broad.
Finally, the compact design of the probe can easily be implemented into a dilution refrigerator, opening a route toward microwave studies on bulk metals and superconductors at temperatures well below 1~K, a temperature range that has rarely been explored for microwave spectroscopy on conducting materials so far.\cite{Lee2001,Bobowski2010,Liu2011,Steinberg2012,Truncik2012}

\section{measurement principle}
The stripline is a transmission line waveguide with TEM wave propagation. The signal runs along a strip-shaped center conductor that is sandwiched between dielectrics and two ground planes (Fig.\ \ref{xsection}). By adding two gaps in this center conductor at a distance $l$ from each other, the traveling wave is partially reflected at these impedance mismatches forming a resonator. The resulting standing wave pattern will allow transmission of electromagnetic waves through the resonator only for frequencies close to the resonance frequencies
\begin{equation}
\nu_{0,n}=\frac{n c}{2\sqrt{\epsilon_r}l}
\label{eq:ResonanceFrequency}
\end{equation}
where $n$ is an integer assigned to the mode number, $c$ the speed of light, and $\epsilon_r$ the permittivity of the dielectric.\cite{pozar1997microwave}  

The frequency dependence of power transmission through this structure can then be restricted to lorentzian curves around the resonance frequencies with bandwidth $BW_n$.\cite{pozar1997microwave} For each mode the quality factor describing the ratio of power stored to the power dissipated per cycle in the resonator can then be defined as $Q_n=2\pi\nu_{0,n}/BW_n$.

As the inverse of the quality factor is essentially a measure of losses, it can be split up into a combination of loss terms
\begin{equation}
  \frac{1}{Q_n}=\frac{1}{Q_e}+\frac{1}{Q_i}=\frac{1}{Q_e}+\frac{1}{Q_d}+\frac{1}{Q_c}.
  \label{eq:Qcombination}
\end{equation}
$Q_e$ is associated to losses through the external coupling, while the internal quality factor $Q_i$ describes the combination of several Q-factors associated to the resonator like $Q_d$, which is defined by losses in the dielectric, and $Q_c$ which is determined by losses in the conductors. For a resonator that is undercoupled to the external circuitry ($Q_e\gg Q_n$) and comprises of low-loss dielectrics ($Q_d\gg Q_n$), the measured quality factor will be solely determined by the losses in the conductors ($Q_c\approx Q_n$).

This is now exploited by using the sample as upper ground plane and superconducting materials for the other conductors. Below the critical temperature $T_c$ of the superconductors, their losses are vanishingly small compared to those of the sample of interest. $Q_c$ will then be dominated by the ohmic losses in the sample, that can be evaluated using the Wheeler inremental inductance rule \cite{Wheeler42formulasfor} and the TEM-nature of the stripline. One then finds that the surface resistance $R_S$ of the sample is connected to the measured values $Q_n$ and $\nu_{0,n}$ by
\begin{equation}
R_S=\Gamma \frac{\nu_{0,n}}{Q_n},
\label{eq:Rs}
\end{equation}

with a constant $\Gamma$ that can be calculated for a known stripline geometry.

\section{measurement setup}
\begin{figure}
  \includegraphics{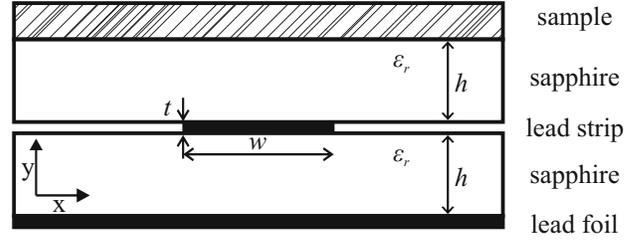}
  \caption{Cross section of the used stripline structure.\label{xsection}}
\end{figure}
The materials of the resonator have to satisfy all the above assumptions required to quantitatively relate the sample surface resistance to the quality factor of the resonance.
As dielectric, sapphire exhibits the required vanishing loss contributions due to its extremely low loss tangent\cite{lowlossdielectric} of ${\tan\delta<10^{-6}}$. Additionally, its high permittivity\cite{lowlossdielectric} between $\epsilon_{r,\bot}=9.3$ and $\epsilon_{r,\parallel}=11.3$ allows small device dimensions without sacrifice in accessible resonance frequency.

For the conductor material, superconductors show the exceptionally low surface resistance needed for this method, however, they are limited by their critical temperature $T_c$. Therefore superconductors with a high $T_c$ are favored, moreover when considering that their surface resistance drops exponentially with temperature. The material chosen in this work was lead, for the relatively high $T_c$ of $7.2$\,K,\cite{kittel1996introduction} the low surface resistance,\cite{PhysRevB.15.4412} and the possibility for simple deposition by thermal evaporation.

The dimensions of the stripline seen in Fig.\ \ref{xsection}, strip width $w$ and thickness $t$, as well as dielectric height $h$ and permittivity $\epsilon_r$, determine the characteristic impedance $Z_c$ of the stripline, which has to be adjusted to the 50\ $\Omega$ impedance of the external circuitry. As no closed-form expression for $Z_c(h,w,t,\epsilon_r)$ has been found to date, either the impedance has to be numerically calculated,\cite{Sheen1991} or one of the many approximation formulas has to be utilized; in the following we will use the one by Wheeler.\cite{wheeler78stripline} 
Some values are given, as $\epsilon_r$ is defined by the sapphire and we fixed the strip thickness $t$ to 1\,\textmu m as a trade-off between exceeding penetration depth and evaporation limitations. To achieve a stripline impedance of 50\,$\Omega$, the Wheeler formula implies a fixed ratio between strip width and substrate height of $w/h\approx0.36$. The formula and calculated impedance values can be found in the appendix. We have chosen sapphire thicknesses of $h=430$\,\textmu m, which has already been successfully applied,\cite{Scheffler2012} as well as a thinner $h=127$\,\textmu m substrate leading to strip widths of $w=155$\,\textmu m and $w=45$\,\textmu m respectively. The smaller dimensions allow a far lower fundamental frequency for a given sample area, and accordingly more frequency points to be measured overall.

Following equation \eqref{eq:ResonanceFrequency}, the fundamental frequency of a stripline resonator is determined by its length. For a sample with a diameter of 5 mm, the fundamental frequency of a straight stripline would be at about 10 GHz. An easy way to reduce this frequency and observe more frequency points given by overtones, is increasing the length of the line by meandering the center conductor beneath the area of the sample.

The center conductor has minimal impedance mismatch if the bends of the meander are shaped by a specially defined right-angle miter or in a curve with large enough radius ${r>3w}$.\cite{wadell1991transmission}
Coupling effects (``crosstalk'') between adjacent parallel sections that would lead to an inhomogenious impedance of the stripline are mitigated by spacing the center strips at least twice the ground plane distance $h$.\cite{coupledstrip}
To avoid radiation losses and interactions with the sample mounting, the strip should also be kept at least at a distance $h$ to the projected edge of the sample.

The desired shape of the strip conductor is thermally evaporated onto the sapphire substrate using shadow masks.
Altering the position and therefore the distance $l$ between two gaps in the center conductor allows tuning the resonance to any frequency upwards from the fundamental frequency given by the maximal achievable length.
Optimizing the gap size $g$ is important as this determines the external quality factor $Q_e$, which in turn has to be far higher than the internal quality factor $Q_i$ to accurately display sample information in the measured value $Q_n$. On the other hand the insertion loss \cite{pozar1997microwave}
\begin{equation}
IL=-20\log(1+Q_e/Q_i)\,\textrm{dB}
\label{eq:IL}
\end{equation}
decreases the detectable signal with increasing $Q_e$. Therefore the gap size has to be adjusted to reach insertion loss values between -20 and -40\,dB as a tradeoff between keeping the systematic error in $Q_n$ below 1\,\% and convenient measurability.

The junction between stripline and external coaxial circuitry is achieved by SubMiniature A (SMA) stripline launchers\cite{Rosenberger} for the thick substrates and smaller sparkplug connectors\cite{Anritsu} for the thinner substrates.

\begin{figure}[t]
  \includegraphics{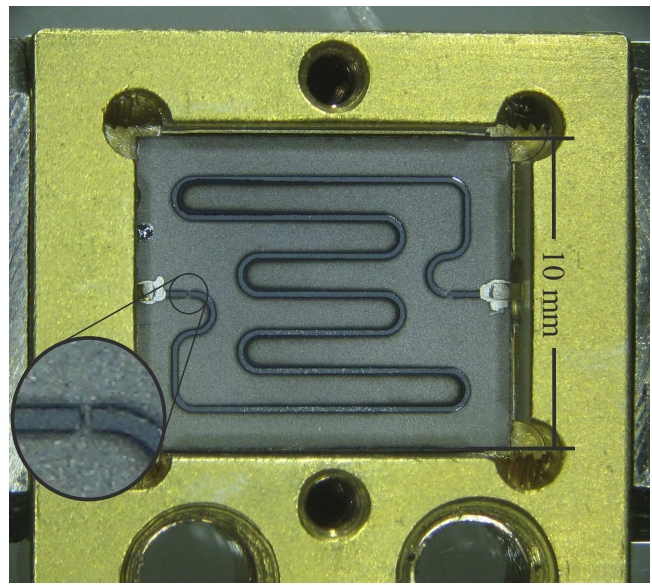}
  \caption{(color online) Photograph of a resonator with a 155\,\textmu m wide center strip in a box designed for the 430\,\textmu m thick sapphire substrates. The meander structure of the inner conductor maximizes the usable length. The gaps in the center conductor are visible right before the first bend next to the launching on the outer sides.\label{meander}}
\end{figure}

To assemble the resonator, the different layers of the stripline are stacked into the brass mounting box. The first is a 8 \textmu m thin lead foil acting as lower ground plane, which is then covered by the sapphire substrate with the evaporated lead strip conductor on top. After this step the stripline launchers are  mounted and connected by silver paint. A resonator at this stage can be seen in Fig.\ \ref{meander}.

Next layer is a bare sapphire with recesses for the used launching. Finally the sample is placed above the area of the resonator. If the sample leaves the stripline between the gap and the launching uncovered, a second lead foil with recesses for the sample and the launching can be placed on the sapphire.

\begin{figure}[t]
  \includegraphics{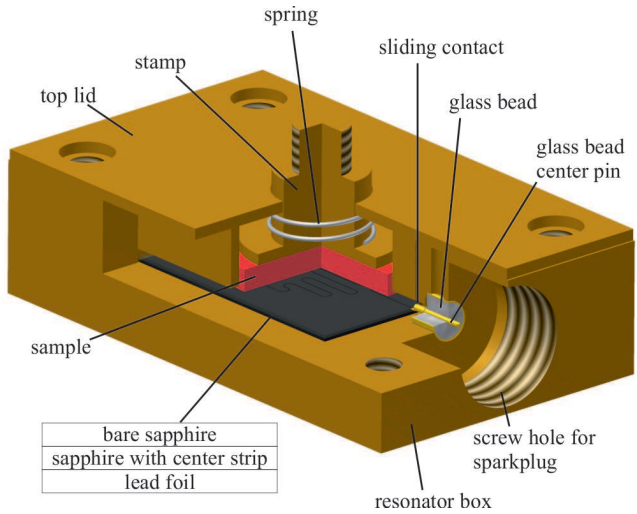}
  \caption{(color online) Design of housing for resonator and sample, including transitions to coaxial cables.\label{box}}
\end{figure}

As geometry and dimensions are crucial for the characteristic impedance of the stripline and therefore for the accuracy of the calculated $\Gamma$-factor, the stripline stack has to be firmly and reliably mounted into the box. To this end, a weak copper-beryllium spring was implemented into the box lid. The spring exerts slight pressure onto a stamp, holding the stripline layers in place.
An illustration of a completely assembled resonator box can be seen in Fig.\ \ref{box}.

The microwave and cryogenic setup used for this work was a slightly modified Corbino spectrometer.\cite{Scheffler2005a}
Here the complex transmission S-parameter S$_{21}$ is measured by a network analyzer (NWA) that is connected to the resonator box via combinations of semi-rigid and flexible coaxial cables.
Additionally a 30dB broadband microwave amplifier ranging from 0.5 to 18~GHz was implemented at the second port of the NWA.\cite{AmplifierMITEQ}

\begin{figure}[t]
  \centering
  	\includegraphics[width=.5\textwidth]{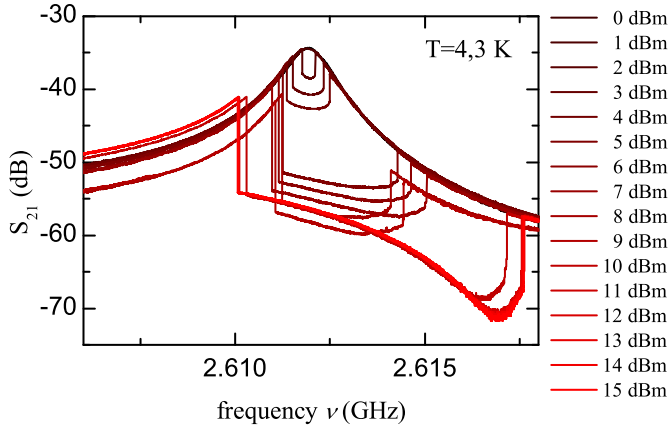}
  \caption{(color online) Power dependence of a resonance mode. The sharp kinks occur when the critical current is exceeded in the center strip. This also leads to heating effects indicated by the frequency shift of the undisturbed mode for high input powers.}
  \label{fig:power_dependence}
\end{figure}

The \textsuperscript{4}He glass cryostat of this setup combined with a rotary and a roots pump allows temperatures at the sample lower than 1.1~K.\cite{Steinberg2008,Steinberg2012}
The mounting for the resonator box holds a temperature sensor and a heater to control temperature.
Before the actual measurement, a broadband spectrum is taken at 7.5~K, slightly above the $T_c$ of the lead resonator, whose phase will later serve in calibration purposes. With lead in the superconducting phase at lower temperatures, the set of observable modes can be identified, which is typically done at liquid helium temperature of $\approx4.3$~K. At this stage one can also perform power-dependent measurements as seen in Fig.\ \ref{fig:power_dependence} to assure that the applied power does not exceed the critical current of the superconducting resonator while at the same time is high enough for convenient signal detection. Breakdowns of superconductivity in the resonator were observed down to powers of -30\,dBm, so typical powers sent into the microwave line were chosen in the range -50\,dBm to -40\,dBm.  After pumping on the helium bath and reaching the lowest temperature, all modes will be measured for each consecutive temperature point. A quick scan of each resonance determines the resonance frequencies $\nu_{0,n}$ and bandwiths $BW_n$, which then allows for detailed resonance measurements centered around the measured mode with a frequency span of 15 times the $BW$.

\section{data analysis}

\begin{figure}
  \includegraphics{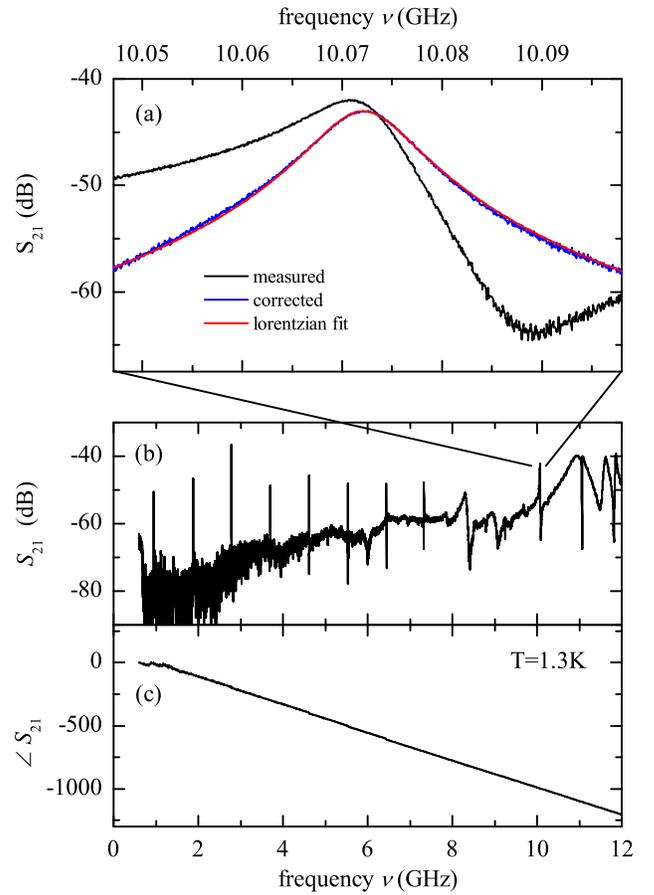}
  \caption{(color online) (a) shows a typical resonance spectrum in detail before and after background correction together with a lorentzian fit to the corrected spectrum. (b) and (c) show the absolute value and the unwrapped phase $\angle$ of the broadband response from the same resonator. Jumps on low frequencies are due to the very low signal, that leads to inconsistencies during unwrapping. The high values of the phase are caused by the long coaxiables cables reaching in and out of the cryostat. \label{background}}
\end{figure}

The measured resonance spectra, e.g. in Fig.\ \ref{background}(a), do not necessarily resemble the anticipated symmetric lorentzians, but can be rather asymmetric. The reason is a residual background $S_{21,bg}$ caused by capacitative transmission from input to output, bypassing the resonator.\cite{Mola2000} This is evident from Fig.\ \ref{background} (b) as the substantial signal at frequencies far away from the narrow resonance peaks. This complex background adds to the symmetric resonance transmission $S_{21,r}$ to give the measured signal
\begin{equation}
|S_{21}|^2=|S_{21,r}|^2+|S_{21,bg}|^2+S_{21,r} S_{21,bg}^*+S_{21,r}^* S_{21,bg}.
\label{eq:measured_signal}
\end{equation}
While a dedicated calibration procedure with measurements of additional reference samples can be used to generically obtain unbiased data, \cite{Ranzani2013,Yeh2013} we follow a different approach and apply a background model
\begin{equation}
S_{21,bg}=r(\nu)\exp\{i(\kappa 2\pi\nu + \Phi)\}
\label{eq:background_model}
\end{equation}
locally for every resonance mode. The model is a combination of the background amplitude $r(\nu)$ as seen in Fig.\ \ref{background} (b), and the phase delay $\kappa$ caused by the microwave cabling seen in figure \ref{background} (c) along with a phase offset $\Phi$. These parameters are then determined for the small spectral range of each resonance spectrum: $r(\nu)$ is assumed constant in the narrow frequency range of the resonance and its value is taken as the mean amplitude at two frequencies well below and above the resonance frequency; $\Phi$ is also taken as the mean of the two respective phases. $\kappa$ is determined by a linear fit on the broadband phase spectrum taken at 7.5~K, in the normal state slightly above $T_c$, where the resonances are not observable over the background. With this model subtracted, the remaining signal resembles the expected lorentzian behavior and an accurate fit of the quality factor is possible. 

To establish the connection between sample surface resistance $R_S$ and the measured values of quality factor $Q_n$ and resonance frequency $\nu_{0,n}$ as stated above in (\ref{eq:Rs}), we first consider the relation of  $Q_n$ to $\nu_{0,n}$ and the attenuation constant $\alpha$: \cite{pozar1997microwave} 
\begin{equation}
Q_n=\pi \nu_{0,n}/\alpha v_p.
\label{Qalpha}
\end{equation}
This is valid for any transmission line resonator with $v_p$ the phase velocity of the electromagnetic wave in the transmission line. The attenuation in the conductors can be calculated by looking at the ohmic losses due to the currents per unit width $J(x)$ present in the skin depth layer of the conductor surfaces \cite{ramo1944fields}
\begin{equation}
\alpha=\frac{R_S}{2Z_c}\int_{-\infty}^\infty\frac{|J(x)|^2}{|I|^2}dx.
\label{eq:AlphaOneSurface}
\end{equation}
Here $Z_c$ is the characteristic impedance of the stripline and $|I|$ gives the magnitude of the total current in the surface.

\begin{figure}
	\centering
		\includegraphics{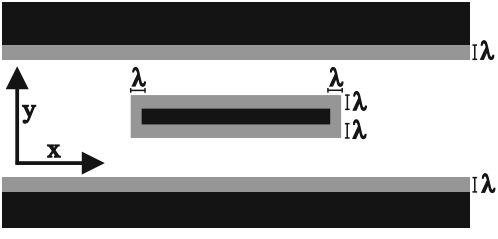}
\caption{Cross section of the stripline with areas penetrated by the electromagnetic wave marked in grey. The Wheeler incremental inductance rule states that the characteristic impedance of this penetrated transmission line is the same as the one of a structure with perfect electric conductors, but surfaces receded by half the characteristic lengthscale of penetration $\lambda$. \label{fig:penetrated_surfaces}}
\end{figure}

The Wheeler incremental inductance rule \cite{Wheeler42formulasfor} states that the characteristic inductivity of a TEM transmission line, that is penetrated by an electromagnetic wave with the characteristic lengthscale $\lambda$, can be calculated as the one of a hypothetical unpenetrated transmission line with surfaces receded by $\lambda/2$ (see Fig.\ \ref{fig:penetrated_surfaces}). The energy of the resulting increase in inductivity $\Delta L$ can then be equated to the magnetic field energy in the penetrated region:
\begin{equation}
\frac{\mu_0}{2}\int_S|H|^2dxdy=\frac{\Delta L |I|^2}{2}.
\label{eq:wheeler1}
\end{equation}
Here $L$ is the inductance given per unit length, the integral is only regarded over the penetrated cross section area $S$.\\
The integration over $|H|^2$ along the axis perpendicular to the surface can be executed using the exponential decay of the field in the conductor. Assuming $\lambda$ is smaller than the thickness of the conductor leads to
\begin{equation}
\frac{\int_{-\infty}^\infty|H|^2dx}{|I|^2}=\frac{2\Delta L}{\lambda\mu_0}.
\label{eq:wheeler2}
\end{equation}
As the magnetic field has the same absolute value and is perpendicular to the surface currents $|H|=|J(x)|$, the integral in \eqref{eq:AlphaOneSurface} can be replaced by \eqref{eq:wheeler2} to find the equation
\begin{equation}
\alpha=\frac{R_S}{Z_c}\frac{\Delta L}{\lambda\mu_0}.
\label{eq:wheeler3}
\end{equation}
By approximating $\Delta L$ with $\frac{\lambda}{2}\frac{\partial L}{\partial y}$ and noting that $L=Z_c/v_p$ for a stripline\cite{pozar1997microwave}, \eqref{eq:wheeler3} can be simplified to
\begin{equation}
\alpha=\frac{R_S}{2\mu_0Z_c}\frac{\partial Z_c}{\partial y}.
\label{eq:wheeler4}
\end{equation}
The connection between the measured quantities quality factor and resonance frequency to the surface resistance is then finally established by inserting \eqref{eq:wheeler4} into \eqref{Qalpha}:
\begin{equation}
R_S=\frac{Z_c}{\frac{\partial Z_c}{\partial y}}2\pi\mu_0 \frac{\nu_{0,n}}{Q_n}=\Gamma\frac{\nu_{0,n}}{Q_n}.
\label{eq:QRs}
\end{equation}
For the characteristic impedance $Z_c$ an approximation formula from Wheeler can be used, that gives $Z_c(h,t,w,\epsilon_r)$ in terms of the stripline geometric properties $h$, $w$, $t$ and dielectric constant $\epsilon_r$. 
I.e.\ $\Gamma$ is a constant and can be calculated for a known geometry. Values of $\Gamma$ calculated for our geometries can be found in the appendix.

\section{measurement results}
\begin{figure}
  \includegraphics{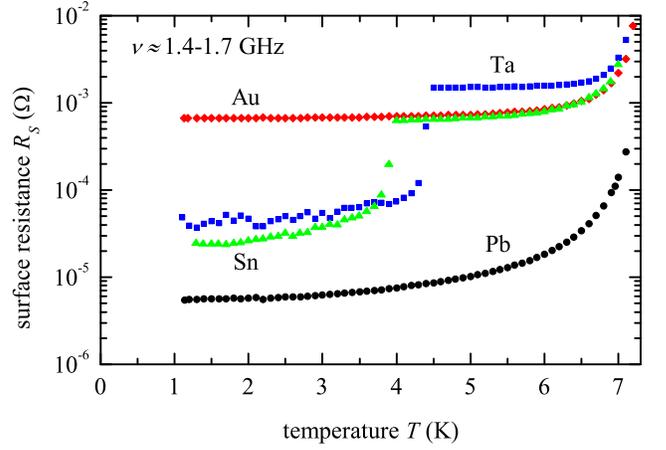}
  \caption{(color online) Temperature-dependent surface resistance as calculated from the measured quality factor and resonance frequency for the materials gold (Au), tantalum (Ta), tin (Sn), and lead (Pb) as top ground plane of a lead resonator. \label{RSvsT}}
\end{figure}

To evaluate the measurement method, various reference materials have been used as sample. Fig. \ref{RSvsT} shows the temperature dependence of the measured surface resistance for the simple metal gold and the elemental superconductors tantalum, tin, and lead. The samples were mostly foils with thicknesses far higher than their respective skin depths, so bulk properties are expected to be measured. The only exception was tin, which was a 1\,\textmu m thin thermally evaporated film. The shown data is measured from the second mode of resonators with a fundamental frequency of approximately 1.5\,GHz. The all lead resonator shows an exponential drop with decreasing temperature to a residual surface resistance as expected for a superconductor.\cite{halbritter:82} The substantial surface resistance of lead close to the critical temperature at 7.2~K also effects the resistance measurements of the other samples for temperatures above 6~K, because the losses in the center strip contribute about 20 times more to the quality factor than losses in a ground plane. At lower temperatures, the losses in the sample materials dominate and reveal the anticipated temperature-independent surface resistance of gold, as well as of tantalum and tin in their normal conducting phase. On the superconducting transitions of the latter two at 4.5\,K and 3.7\,K respectively, the same characteristic exponential drop to a residual value as in the case of lead indicates the BCS-behavior of these elements.

\begin{figure}
  \includegraphics{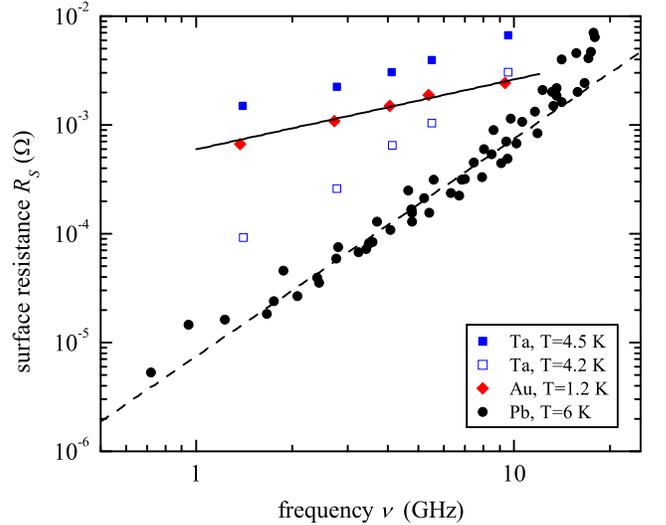}
  \caption{(color online) Surface-resistance data versus frequency measured with one resonator for gold (open diamonds) at 1.2\,K as well as tantalum above (black squares) and below (open squares) the superconducting phase transition. Lead data (black circles) was taken at 6\,K using eight different resonators. The solid line is a power-law fit to the gold data giving an exponent of 0.64. The dashed line represents a power dependence of $R_S\propto\nu^2$ epxected for a superconductor with low photon energies.\cite{Dressel2002a} \label{RSvsf}}
\end{figure}

\begin{figure}
  \includegraphics{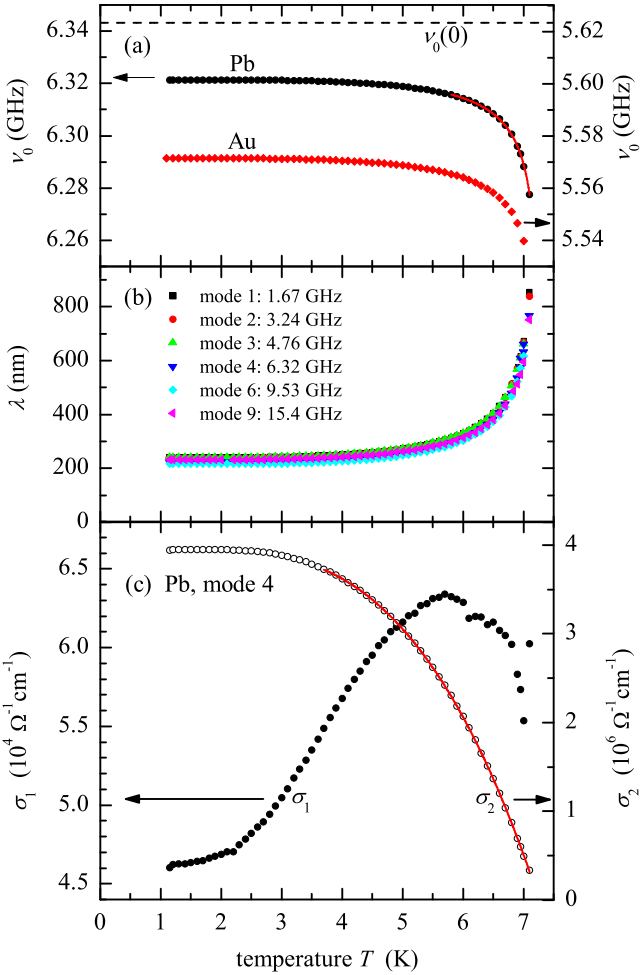}
  \caption{(color online) (a) shows the measured resonance frequencies $\nu_0$ against temperature for one mode of an all lead resonator (black circles) as well as the undisturbed resonance frequency $\nu_0(0)$ as obtained from the fit represented by the dashed line. The open diamonds represents the resonance frequencies of a comparable mode for a resonator with gold replacing a top plane. $\nu_0(0)$ together with the measured frequencies for lead can be used to calculate the penetration depths $\lambda$ that are displayed in (b) for all modes. $\lambda$, in combination with the surface resistance, gives the complex conductivity plotted in (c). The real part $\sigma_1$ shows a noticable coherence peak at around 6\,K, while the imaginary part $\sigma_2$ can be used to fit the superconducting energy gap (solid line). \label{coherencepeak}}
\end{figure}

The frequency dependence of the surface resistance as obtained by measuring multiple modes of the resonator is plotted in Fig.\ \ref{RSvsf} for gold, tantalum, and lead.
A power-law fit to the gold data shows a frequency dependence $R_S\propto\nu^{0.64}$. This suggests the metal is in the anomalous skin effect regime, which is characterized by $R_S\propto\nu^{2/3}$ and no temperature dependence at low temperatures.\cite{Dressel2002a}
The latter can be seen in Fig. \ref{RSvsT}, with the gold surface resistance being constant below 5\,K. Fig. \ref{coherencepeak}(a) further consolidates this assumption, as no change in resonance frequency is seen on lower temperatures for a resonator loaded with gold. As will be seen later on, the resonance frequency is linked to the penetration depth, so a constant frequency denotes no change in skin depth, which is consistent with the anomalous regime.\cite{Dressel2002a}

Fig.\ \ref{RSvsf} also shows that lead follows $R_S\propto\nu^{2}$ as expected for a superconductor with photon energies smaller than the superconducting gap.\cite{Dressel2002a}
At the highest frequencies deviations from the quadratic behavior for lead are visible. These might stem from undesired resonance modes in the resonator box and are not intrinsic to the sample. This leads to a lower $Q$-factor and therefore higher values for the determined surface resistance.
Fig.\ \ref{RSvsf} further depicts data for tantalum at two temperatures: in the normal conducting phase slightly above the critical temperature, where we again find behavior consistent with the anomalous skin effect, and in the superconducting phase slightly below.
Upon this transition the surface resistance drops more than a decade over a temperature range of 0.3\,K and the frequency dependence changes visibly toward $R_S\propto\nu^2$ as seen for lead.

For the special case of a resonator with all conductors made from the same material, in our case lead, it is possible to retrieve its complete optical function by also taking into account the resonance frequency change.
Fig.\ \ref{coherencepeak}(a) shows the resonance frequency as a function of temperature. The increase in resonance frequency when lowering the temperature originates from the decrease of the penetration depth in the surfaces of the stripline. This can be evaluated using that the resonance frequency in a transmission line resonator is related to its characteristic inductance:\cite{pozar1997microwave} $\nu_0\propto \frac{1}{\sqrt{L}}$. As $L$ is also proportional to the characteristic impedance $Z_c$ in a TEM-structure, the resonance frequency of a penetrated stripline can be linked to the penetration depth of the conductor materials by 
\begin{equation}
\nu_0(\lambda)=\nu_0(0)\sqrt{\frac{Z_c(0)}{Z_c(\lambda)}}.
\label{nu_lambda}
\end{equation}
$\nu_0(\lambda)$ and $Z_c(\lambda)$ are the resonance frequency and the characteristic impedance for the penetrated structure, while $\nu_0(0)$ and $Z_c(0)$ are the corresponding values for an unpenetrated structure with perfect conductors.
Similar as above, the change in impedance upon penetration can be approximated by using the Wheeler incremental inductance rule to find
\begin{equation}
\nu_0(\lambda)=\nu_0(0)\sqrt{\frac{1}{1+ \frac{\lambda\pi\mu_0}{\Gamma}}}.
\label{nu_lambda_gamma}
\end{equation}
Because we look at the penetration in all conductor surfaces of the stripline, $\Gamma$ is here a sum of all $\Gamma$-factors for their respective penetrated surfaces as depicted in Fig. \ref{fig:penetrated_surfaces}.
In order to calculate $\lambda$ from the measured resonance frequencies, the undisturbed resonance frequency has to be found, which can be done by measuring a reference material with known penetration depth for the exact same resonator. For a superconductor it is possible to fit this value by inserting the charateristic penetration depth change\cite{Tinkham1996}
\begin{equation}
\lambda=\frac{\lambda(0)}{\sqrt{1-(T/T_c)^4}}
\label{lambda_sc}
\end{equation}
in the temperature range $0.8T_c<T<T_c$ into equation (\ref{nu_lambda_gamma}). The resulting function can be used to fit the measured values with the resonance frequency $\nu_0(0)$ of an unpenetrated stripline, the zero-temperature penetration depth $\lambda(0)$, and the critical temperature $T_c$ as fit parameters, as seen in Fig.\ \ref{coherencepeak}(a).
Now the undisturbed resonance frequency can be used to calculate the penetration depth for all temperatures by solving
\eqref{nu_lambda} for $\lambda$:
\begin{equation}
\lambda=\frac{\Gamma}{\pi\mu_0}\left[\left(\frac{\nu_0(0)}{\nu_0(\lambda)}-1\right)^2\right].
\label{eq:lambda-from-resfreq}
\end{equation}
The resulting penetration depths for all observed modes for this resonator are plotted in Fig.\ \ref{coherencepeak}(b). All modes show the same penetration depth as expected.\cite{Dressel2002a}
These values can be further used to calculate the reactance of the lead films by using the linear relation between reactance $-X_S$ and $\lambda$ for a superconductor:\cite{Dressel2002a}
\begin{equation}
X_S=-377\Omega\frac{2\pi\nu}{c}\lambda
\label{eq:XS}
\end{equation}

which is valid for frequencies $h\nu\ll\Delta$, where $h$ is the Planck constant. This is given as our resonator frequencies are far lower than the energy gap $\Delta$ in lead, as will be seen later on.
Now the surface reactance $X_S$ together with the surface resistance $R_S$, as obtained from the measured quality factor, give a complete set of optical constants in the form of the surface impedance $\hat{Z}_S=R_S+iX_S$. This can also be converted into the complex conductivity $\hat{\sigma}=\sigma_1+i\sigma_2$ that is often used in optics.\cite{Dressel2002a}
The resulting real part of the conductivity is plotted in \ref{coherencepeak}(c). It shows a coherence peak below the transition temperature at around 6 K and then drops to a finite value. The first time a coherence peak in $\sigma_1$ has been measured for lead was by K. Holczer \textit{et al.}\cite{Holczer1991875} The position of their observed peak resembles the measurement presented here, however there are slight devations in the residual conductivity $\sigma_1(T=0)$. This is caused by a differing residual surface resistance $R_S$ that leads to a different residual conductivity at $T=0$\,K. \\
The resulting imaginary part of the conductivity can be seen in Fig.\ \ref{coherencepeak}(c). It rises with decreasing temperature until a final value is reached. A simple analytical function for the temperature dependence of $\sigma_2$ can be found using \cite{Dressel2002a}
\begin{equation}
\sigma_2(T)\approx \frac{A\pi}{h}\\\Delta(T) \cdot \tanh\left(\frac{\Delta(T)}{2k_BT}\right),
\label{sigma2}
\end{equation}
together with the second order taylor expansion for the energy gap\cite{Ferrell1964}
\begin{equation}
\Delta(T)=\Delta(0) \sqrt{3.016\left(1-\frac{T}{T_c}\right)-2.4\left(1-\frac{T}{T_c}\right)^2},
\label{Delta}
\end{equation}
valid for the temperature range $T>T_c/2$. This function can now be fitted to the conductivity data with a parameter $A$, the zero-temperature energy gap $\Delta(0)$ and critical temperature $T_c$.

By analyzing all the observed modes we find the mean value for the zero temperature gap $\Delta(0)=1.59$ meV and the mean transition temperature of $T_c=7.24$ K. This leads to a value of
\begin{equation}
\frac{2\Delta(0)}{k_BT_c}=5.3\pm0.8,
\label{eq:deltaresult}
\end{equation}
which is considerably higher than the theoretical weak-coupling BCS-value of $\frac{2\Delta(0)}{k_BT_c}=3.52$, which indicates the strong-coupling behavior in superconducting lead in consistency with previous works that found $\frac{2\Delta(0)}{k_BT_c}=4.7$ for bulk lead.\cite{anderson1989physicist} The experimental error for the superconducting gap is given by an error margin in the absolute value of the surface resistance and the penetration depth introduced by an uncertainty in the $\Gamma$ factor, which translates directly to an error of the absolute value for the superconducting gap. The fitted transition temperature is not effected by this.

\section{Evaluation and outlook}

The only source of statistical error in this measurement method is the negligible electric noise measured by the NWA.
The biggest systematical error for this measurement is introduced by deviations mainly in the geometric values $w$ and $h$. $\Delta w=10$\,\textmu m is possible when the strip is evaporated using a shadow mask, while $\Delta h=5$\,\textmu m has to be assumed due to irregularities in thickness and spaces in assembly. These lead to uncertainties of the geometric factor $\Gamma$
\begin{equation}
\frac{\Delta\Gamma}{\Gamma}=\frac{1}{\Gamma}\left(\left|\frac{\partial \Gamma}{\partial w}\right|\Delta w + \left|\frac{\partial \Gamma}{\partial h}\right|\Delta h\right)=11\%
\label{eq:relative-error}
\end{equation}
for $w=45$\,\textmu m and $h=127$\,\textmu m.
Because the $\Gamma$ factor is directly proportional to the surface resistance (eq. \eqref{eq:Rs}), the error in $\Gamma$ also translates directly to an error in the surface resistance.

It is important to note that this error margin
only affects the absolute value of the surface resistance. The behavior as a function of temperature is unaffected, because the geometry stays constant. Concerning the frequency dependence, however, the geometry factor can in principle vary, as each mode probes different resonator regions due to the differences in the standing wave patterns.

With our test measurements we have demonstrated how stripline resonators provide a convenient experimental access to frequency-dependent information about the surface resistance of bulk samples at cryogenic temperatures. With its good sensitivity, the applicability to generic bulk samples, and its ability to measure several resonance frequencies at the same time, it might serve as a powerful addition to the established microwave spectroscopy techniques. The stripline resonator shares these main advantages with another, recently established approach for low-temperature surface-impedance measurements that is based on dielectric resonators.\cite{Huttema2006,Truncik2012} Both techniques are well-suited for studies of highly conductive bulk samples, but there are also differences: the dielectric resonator is not fundamentally limited toward higher temperatures, whereas $T_c$ of our resonators naturally sets an upper bound. Similarly, dielectric resonators are well-suited for applications in high magnetic field. Superconducting resonators such as ours, on the other hand, strongly depend on magnetic field,\cite{Scheffler2013} but strategies to mitigate these effects are currently being investigated.\cite{Bothner2012} One advantage of the stripline resonators is their simple fabrication, which basically means just one planar structure on a chip: if a particular probing frequency is desired, the corresponding resonator can easily be prepared by adjusting the length of the meandering line. Furthermore, anisotropic properties could be probed easily by rotating the sample on top of the parallel lines of the resonator. Finally, the accessible frequencies can be very low if the sample is big enough: in our case, typical fundamental frequencies could be 1~GHz or lower, and these values could be decreased even further (using thinner dielectrics) while maintaining the small overall size of the resonator which assures straightforward compatibility with operation in a dilution refrigerator.

One route to improve the performance of our superconducting stripline resonators in the future is switching from evaporated lead films to a different superconductor: an obvious choice would be sputtered niobium films and lithographic patterning.\cite{DiIorio1988} The higher transition temperature $T_c$ = 9.3\,K of niobium \cite{ashcroft1976solid} would directly lead to a higher accessible temperature. Furthermore this also leads to a smaller surface resistance contribution from the BCS part,\cite{Dressel2002a,DiIorio1988} and the surface effects leading to residual resistance of Nb are far better studied than those of lead because of the extensive application of niobium in accelerator technology.\cite{padamsee2008rf} Candidate materials with even higher $T_c$ are NbN and NbTiN,\cite{Oates1991,Ranjan2013} and these feature also higher critical magnetic field, which is required for the study of magnetic-field induced phase transitions.\cite{Scheffler2013}

\section{Conclusion}
We demonstrated a successful application of superconducting stripline resonators for measuring the surface resistance of conductive bulk samples. Using equidistant resonance modes we were able to retrieve frequency-dependent data at cryogenic temperatures for frequencies from 0.5 to 20\,GHz.
We further presented a correction method that allowed us to achieve these data in a single run without the need for reference measurements. For stripline resonators fabricated from a single superconducting material we also showed a method to retrieve a complete set of optical constants for every resonance mode. This allowed us to observe the coherence peak and to determine the superconducting energy gap of lead.

We are currently looking forward to applying these methods for studies of materials that can only be synthesized as bulk samples such as most heavy-fermion materials or unconventional superconductors.\cite{Scheffler2013} Many of these materials display features worth taking a closer look in this frequency and temperature region, making our method a useful tool in an active field of present research.

\begin{acknowledgments}
We thank Gabriele Untereiner for resonator and sample fabrication. Financial support by the DFG is thankfully acknowledged.
\end{acknowledgments}

\appendix
\section{Calculation of the characteristic impedance $Z_c$ and the $\Gamma$-factors}
Wheeler gives the approximate impedance for a stripline with infinitely thin center strip and parameters substrate height $h$, zero-thickness strip width $w'$, and dielectric permittivity $\epsilon_1$ as\cite{wheeler78stripline}
\begin{equation}
Z_c=\frac{30}{\sqrt{\epsilon_1}}\ln\left\{1+\frac{1}{2}\frac{16h}{\pi w'}\left(\frac{16h}{\pi w'}+\sqrt{\left(\frac{16h}{\pi w'}\right)^2+6.27}\right)\right\}.
\label{eq1}
\end{equation}
For finite strip thickness the equivalent zero-thickness strip width $w'$ can be evaluated by\cite{wheeler78stripline}
\begin{equation}
w'=w+\frac{t}{\pi}\ln\frac{e}{\sqrt{\left[\frac{1}{3h/t+1}\right]^2+\left[\frac{1/4\pi}{w/t+1.10}\right]^\frac{6}{3+t/h}}},
\label{eq2}
\end{equation}
where $w$ is the real strip width and $t$ the center strip thickness.\\
The calculated values for the two different geometries we used can be found in table \ref{table:geometry}. An estimation of $\epsilon_1=10$ is used for the sapphire permittivity that can lead to small deviations in the values for $Z_c$. However, this does not affect the resulting values for the surface resistance as the permittivity is canceled out in the expression for the $\Gamma$-factor.

\begin{table}[h]
\centering
	\begin{tabular}{|c|c|c|c|c|c|c|c|}
	\hline
		 $t$ in \textmu m & $w$ in \textmu m & $\epsilon_1$ & $h$ in \textmu m & $Z_c$ in $\Omega$ & $\Gamma$ in $\Omega/\textrm{GHz}$ & $\Gamma_a$ in $\Omega/\textrm{GHz}$\\
		\hline
		 1 & 155 & 10 & 430 & 50.06 & 9.09 & 0.456 \\
		\hline
		 1 & 45 & 10 & 127 & 49.79 & 2.68 & 0.151 \\
		\hline
	\end{tabular}
	\caption{Properties of the two used resonator geometries and their calculated characteristic impedance $Z_c$ and their $\Gamma$-factors}
\label{table:geometry}
\end{table}

As derived from the Wheeler incremental inductance rule, the $\Gamma$-factor for a superconducting stripline where the sample substitutes one ground plane is calculated from the characteristic impedance, its derivative in direction of skin layer penetration, and the magnetic constant $\mu_0$:
\begin{equation}
\Gamma=\frac{Z_c}{\frac{\partial Z_c}{\partial h}} 2\pi\mu_0.
\label{eq3}
\end{equation}
In our case the penetration can be seen as an effective increase in substrate height, so the derivative is taken in respect of $h$.\\
For a stripline where all conductors consist of the same material, losses in all their surfaces have to be considered. Following the concept of receding surfaces, this translates to an increase in $h$ for both ground planes, as well as a decrease in $w$ and $t$ from both sides of the center strip. With all these parts considered in the attenuation constant this leads to a different $\Gamma$-factor
\begin{equation}
\Gamma_a=\frac{Z_c}{\frac{\partial Z_c}{\partial h}-\frac{\partial Z_c}{\partial w}-\frac{\partial Z_c}{\partial t}}\pi\mu_0.
\label{eq4}
\end{equation}
The resulting $\Gamma$-factor values for each of the two used stripline geometries are also displayed in table \ref{table:geometry}.

\end{document}